\documentclass[12pt]{article}
\usepackage{graphicx}

\begin{document}

\title{All-spherical telescope \\ with extremely wide field of view}
\author{V.~Yu.~Terebizh$^{1,2}$\thanks{E-mail: valery@terebizh.ru}\\
$^{1}$Crimean Astrophysical Observatory, Nauchny, Crimea 298409\\
$^{2}$Institute of Astronomy RAN, Moscow 119017, Russian Federation}

\date{February 07, 2016}

\maketitle

\begin{abstract}
An all-spherical catadioptic telescope with the angular field of view 
of several tens of degrees in diameter and spherical focal surface is proposed 
for the monitoring of large sky areas. We provide a few examples of such a system 
with the apertures up to $800$~mm and the field of view $30^\circ$ and $40^\circ$ 
in diameter. The curvature of the focal surface is repaid by high performance of 
the telescope. In particular, the diameter of a circle, that includes 80\% of 
energy in the polychromatic image of a star, is in the range $1.4'' - 1.9''$ 
across the field of $30^\circ$ size and $2.2'' - 2.9''$ for the field of 
$40^\circ$ size. Some ways of working with curved focal surfaces are discussed.
\end{abstract}

\section{Introduction} 

There are two primary modes in surveying of large areas of the sky:
1)~we need to cover {\em sequentially} the area in the reasonable time; 
2)~the sky area we are interested in should be under {\em continuous} 
watching, as is the case when we look for the fast transient objects.  
To some extent, problems of the first kind can be solved with the help of 
wide-field telescopes with a flat field of view, such as those discussed by
Wilson (1996) and Terebizh (2011). It is easily seen that problems of the 
second kind require too many flat-field telescopes, so it seems that the 
better way in this case is the creation of a single telescope with an 
extremely wide angular field, even at a spherical focal surface. 

\begin{figure}   % Fig. 1
   \centering
   \includegraphics[width=80mm]{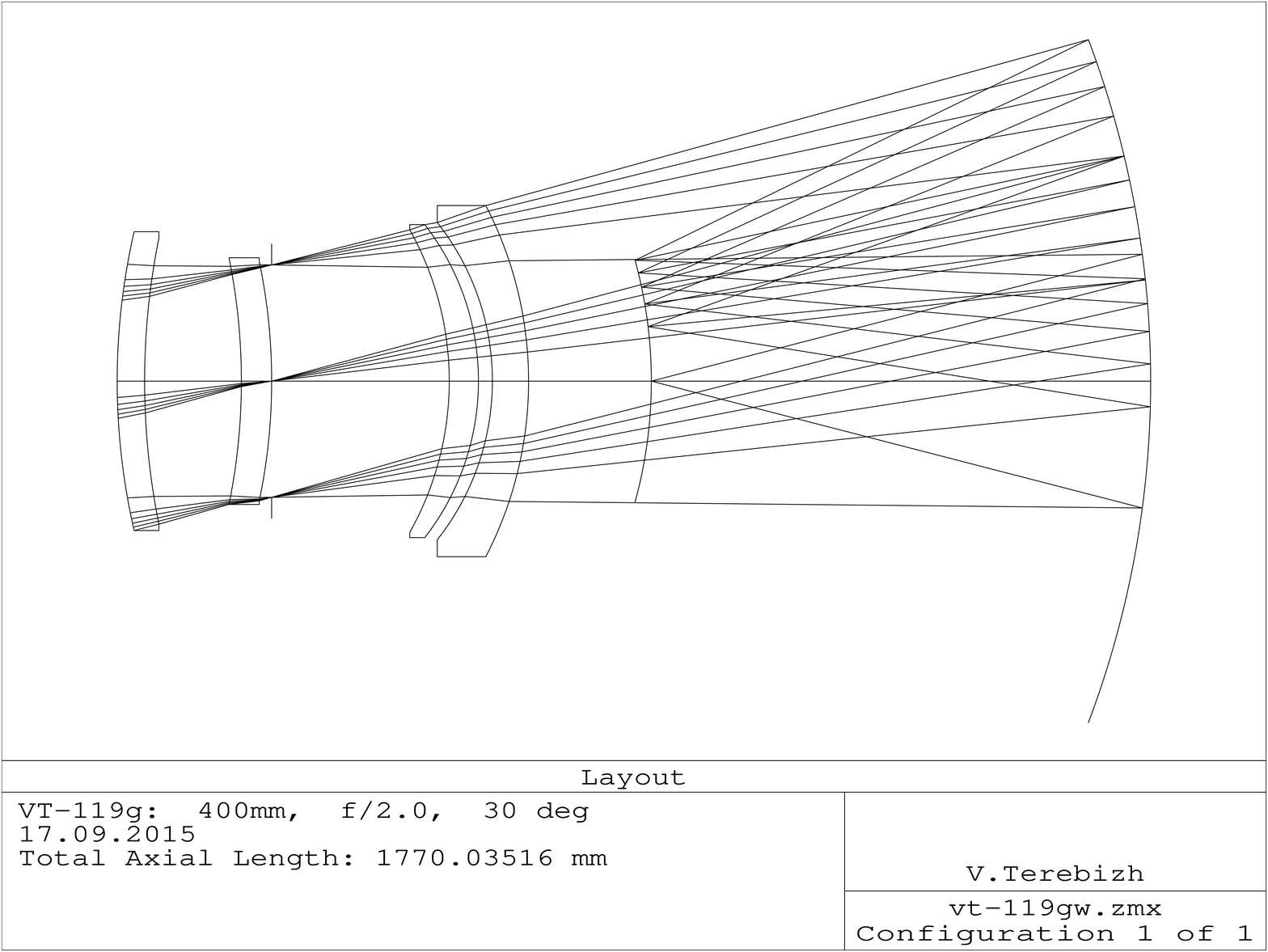}
   \caption{Optical layout of design VT-119g with entrance pupil diameter 
   of $400$~mm and a $30^\circ$ field.}
\end{figure}

Just this way was chosen in the second half of the last century, when the 
Baker-Nunn (Henize 1957; Baker 1962), Hawkins and Linfoot (1945), and 
Maksutov-Sosnina (1950s, unpublished) cameras were put into operation. Their 
angular field attained $20^\circ-30^\circ$, whereas the shielding of light and 
curvature of the focal surface were taken into account by application of narrow 
emulsion tape. The main disadvantages of these cameras were as follows: 
1)~some lens surfaces were substantially aspheric; 2)~the demanding sorts of 
glass were used in the correctors; 3)~nevertheless, the image quality was 
inadequate. For example, four surfaces of the Baker-Nunn camera were aspheres 
of 4th and 8th orders, the Schott KzFS2 and SK14 glasses were applied, but the 
rated image of a point-like source of light was nearly $100\,\mu$m ($40''$) in 
diameter (Carter et al. 1992). Later modification of the Baker-Nunn camera with 
aspheric surfaces of the same order but larger aperture provides 1 arc minite 
resolution in a narrow spectral band (Sasaki et al. 2002).

\begin{figure}   % Fig. 2
  \centering
  \includegraphics[width=80mm]{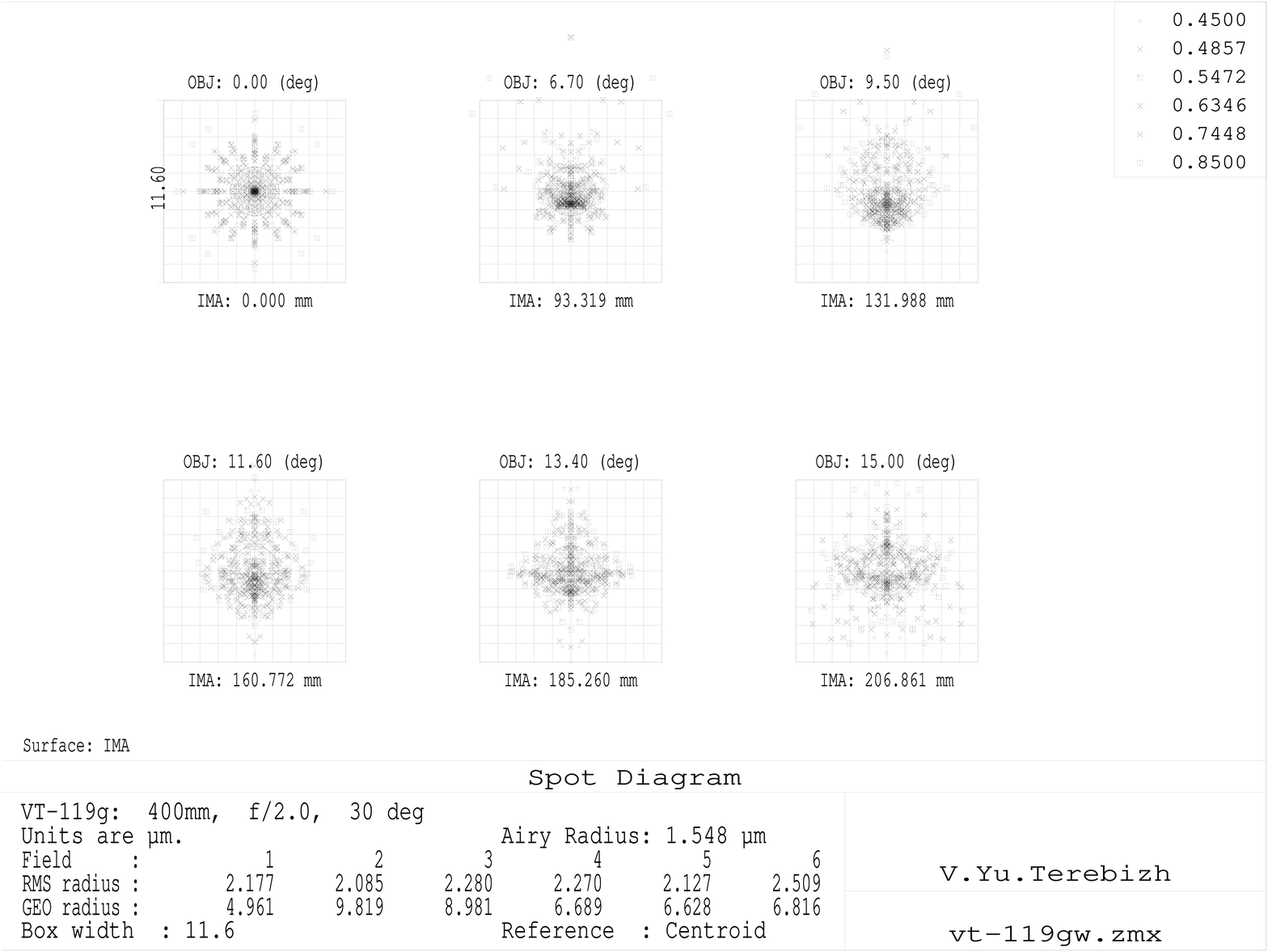} 
  \caption{Spot diagram of VT-119g in the polychromatic waveband $0.45-0.85\,\mu$m.
     The field angles are $0^\circ$, $6.7^\circ$, $9.5^\circ$, $11.6^\circ$, 
     $13.4^\circ$ and $15.0^\circ$ (equal areas). Airy disc diameter is 
     $3.1\,\mu$m, box width is $11.6\,\mu$m $\simeq 3''$.}
\end{figure}

In view of the above, it was somewhat surprisingly that an {\em all-spherical 
system made of simplest types of glass} yet provides nearly diffraction-limited 
polychromatic images in the field of the order of $30^\circ$ in diameter 
(Terebizh 2015). A few further examples of such a system are given 
below\footnote{In calculations, we used the {\em Zemax} optical program 
(ZEMAX Development Corporation, U.S.A.).}.

\section{All-spherical telescope with a $30^\circ$ field} 

An example shown in Fig.~1 has been designed for the aperture $400$~mm, effective 
focal length $800$~mm ($f/2.0$), waveband $0.45-0.85\,\mu$m and the $30^\circ$ 
field of view. All lenses can be made of the same material; we prefer the fused 
silica because of its stability and excellent optical properties, in particular, 
high UV-transparency. 

As one can see from Fig.~2, the image quality is nearly constant across the field. 
More specifically, the $D_{80}$~-- diameter of a circle, that includes 80\% of 
energy in the polychromatic image of a star, varies from $5.4\,\mu$m ($1.4''$) on 
the optical axis to $7.3\,\mu$m ($1.9''$) on the edge of the field. The complete 
description of the system is given in Table~1. 

The proposed system proceeds from the two generic versions of the Bernhard 
Schmidt (1931) camera, that were then developed by A.~Bouwers (1941, 1946), 
D.D.~Maksutov (1944), D.G.~Hawkins and E.H.~Linfoot (1945), C.G. Wynne (1947) 
and J.G.~Baker (1945, 1962)\footnote{See Wachmann (1955); Busch, Ceragioli and 
Stephani (2013) for historical roots, Wilson (1996); Rutten and Venrooij (1999); 
Schroeder (2000) and Terebizh (2011) for specifications and discussion.}. Our 
current goal is to get rid completely of aspheric surfaces. This is partly achieved 
by introduction the modified double meniscus of Wynne (1947), i.e., the first and 
forth lenses in Fig.~1; the two inner lenses were inserted both to minimize coma 
and spherochromatic aberration\footnote{The system with only three lenses provides 
noticeably worse images than the discussed one, while the increasing the number of 
lenses slightly improves images, but seems too massive.}. The double-meniscus 
corrector was applied by Baker in his super-Schmidt design, but he placed inside 
a highly aspheric correction plate such as that introduced by Schmidt. Meanwhile, 
the only possibility to achieve the true point symmetry about the center of the 
entrance pupil is to use a purely spherical optics. 

% -------------------- Table 1
\begin{table}
\caption{VT-119g design with an aperture of 400~mm and $30^\circ$ field of view.
The effective focal length is 800~mm.}
\begin{tabular}{cccccc}
\hline \noalign{\smallskip} 
Surf.& Com-   & $R_0$       & $T$         & Glass   & $D$    \\
 No. & ments  & (mm)        & (mm)        &         & (mm)   \\
\hline 
 1   & L1     & 1141.913    & 47.0        & FS      & 511.8  \\
 2   &        & 1241.285    & 165.714     & --      & 490.1  \\
 3   & L2     & $-$1060.80  & 51.871      & FS      & 422.5  \\
 4   &        & $-$997.351  & 0.0         & --      & 410.1  \\
 5   & Stop   & $\infty$    & 304.135     & --      & 397.9  \\
 6   & L3     & $-$530.106  & 50.0        & FS      & 518.3  \\
 7   &        & $-$436.465  & 23.871      & --      & 536.0  \\
 8   & L4     & $-$438.102  & 62.0        & FS      & 543.4  \\
 9   &        & $-$653.243  & 1065.44     & --      & 601.2  \\
10   & M1     & $-$1656.21  & $-$855.191  & Mirror  & 1169.6 \\
11   & Image  & $-$781.790  & --          & --      & 413.8  \\
\noalign{\smallskip}\hline
\end{tabular} 

Notes to Table~1:\\
$R_0$~-- paraxial curvature radius, $T$~-- distance to the next surface, 
$D$~-- diameter, FS~-- fused silica. All surfaces are spheres. 
\end{table}

The point symmetry both of the fundamental form of a wide-field telescope 
consisting of an idle stop at the center of curvature of a spherical mirror 
(J.~Petzval, H.~Vogel, K.~Strehl) and the basic Schmidt's model is closely related 
to the fact that the entrance pupil coincides with the aperture stop. To a good 
approximation, the same is also true for our design. 

\begin{figure}   % Fig. 3
  \centering
  \includegraphics[width=80mm]{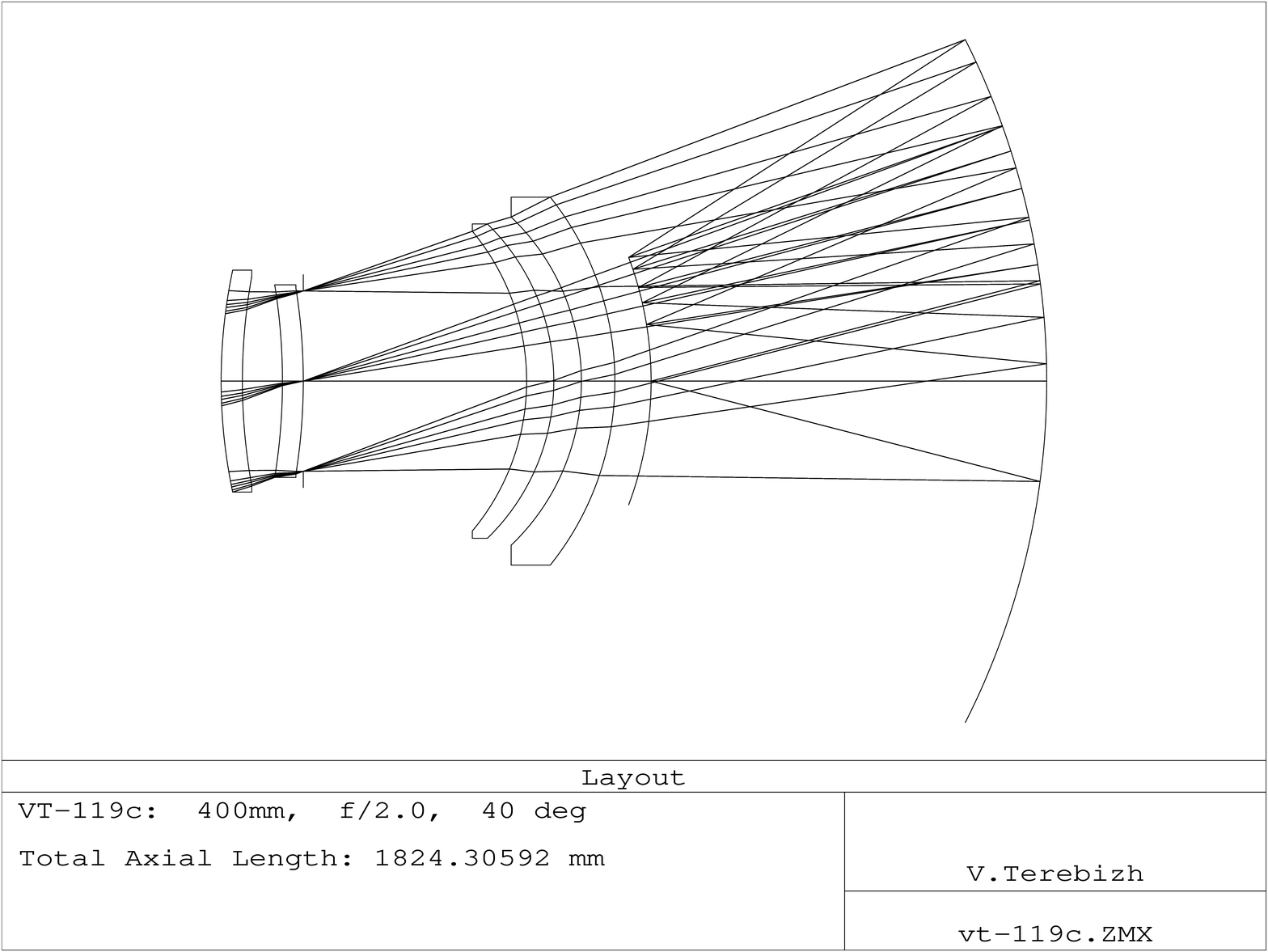} 
  \caption{Optical layout of design VT-119c with entrance pupil diameter of 
    $400$~mm and a $40^\circ$ field.}
\end{figure}

Another its important feature is that the optical power of the system is provided 
by the mirror, while the four-lens corrector working at $f/44$ is nearly afocal. 
Just that should be the case to prevent the chromaticity at reducing spherical 
aberration inherent to the spherical mirror. The resulting width of the chromatic 
focal shift curve is $10.5\,\mu$m, whereas it should be less than $\sim 9\,\mu$m for 
an ideal, diffraction-limited system. Besides, a small optical power of the lens 
corrector greatly contributes~-- along with the spherical shape of surfaces~-- to 
the mitigation of general tolerances. 

As regards losses of light, a strip-like detector of size 
$30^\circ \times 5^\circ$ ($39$~cm~$\times\,6.5$~cm) shields less than 7\% of flux. 

Obviously, individual light detectors may be arranged freely, both continuously 
and discretely. An optimum way is to arrange them in accordance with the shape of 
the observed sky area.

\section{All-spherical telescope with a $40^\circ$ field} 

To illustrate the capabilities of the optical layout under discussion, 
we give in Fig.~3 and Table~2 an example of a $40^\circ$-design VT-119c, which 
spherical lenses are made of fused silica and simplest glasses Schott N-F2 and 
N-BK7. A similar system that uses only fused silica provides just a little 
inferior image quality, but is more expensive. 

% -------------------- Table 2
\begin{table}
\caption{VT-119c design with an aperture of 400~mm and a $40^\circ$ field.
The effective focal length is 800~mm.}
\begin{tabular}{cccccc}
\hline \noalign{\smallskip} 
Surf.& Com-   & $R_0$       & $T$         & Glass   & $D$    \\
 No. & ments  & (mm)        & (mm)        &         & (mm)   \\
\hline 
 1   & L1     & 1190.292    & 47.000      & FS      & 490.4  \\
 2   &        & 1305.352    & 88.531      & --      & 463.9  \\
 3   & L2     & $-$1326.54  & 46.000      & N-F2    & 425.5  \\ 
 4   &        & $-$1300.49  & 0.000       & --      & 410.8  \\
 5   & Stop   & $\infty$    & 493.552     & --      & 398.5  \\
 6   & L3     & $-$516.800  & 60.000      & FS      & 662.2  \\
 7   &        & $-$484.196  & 60.969      & --      & 694.6  \\
 8   & L4     & $-$501.003  & 74.000      & N-BK7   & 725.2  \\
 9   &        & $-$649.809  & 954.254     & --      & 813.0  \\
10   & M1     & $-$1671.01  & $-$874.182  & Mirror  & 1509.0 \\
11   & Image  & $-$777.519  & --          & --      & 546.7  \\
\noalign{\smallskip}\hline
\end{tabular} 

Notes to Table~2:\\
Designations are the same as in Table~1. The glasses N-F2 and N-BK7
are from the Schott sample. All surfaces are spheres.
\end{table}

The effective focal length of VT-119c is still equal to $800$~mm, the design 
waveband remained $0.45-0.85\,\mu$m. As before, the image quality is 
nearly fixed across the field: the diameter $D_{80}$ of a star image varies 
in the range $8.5-11.1\,\mu$m ($2.2''-2.9''$). The width of the chromatic 
curve, $27.8\,\mu$m, is not far from that for the diffraction-limited system, 
$10.2\,\mu$m.

\section{Survey speed and limiting magnitude} 

When working with wide-field telescopes, most important characteristics are
the survey speed $S$\,(degree$^2$/second) and limiting magnitude $m_{lim}$ 
given the exposure time $T$. Let us consider in this context the VT-119g 
design with the rectangular field of view of, say, $30^\circ \times 5^\circ$, 
giving the observed sky area $\Omega = 150$ deg$^2$ (the diameter of the 
equivalent circular field  $2w_e \simeq 13.8^\circ$). 
It was assumed that the noise obeys the Poisson distribution.
Assuming also that the telescope image quality $D_{80} = 1.7''$, its 
transparency is $0.85$, the fraction of unvignetted rays $U = 0.90$, the 
quantum efficiency of detector is $0.85$ counts/photon, pixel size is 
$9\,\mu$m, the atmosphere seeing $\beta_{atm} = 1.5''$, the sky background 
is $20.0^m$/arcsec$^2$, the optical thickness of the atmosphere in zenith 
is $0.30$, the object zenith angle is $40^\circ$, the dead time\footnote{The 
`dead time' is the sum of read-out and telescope redirection time spans.} 
is $5$~sec, and the threshold signal-to-noise ratio $S/N = 8$, we come to 
results shown in Fig.~4. Our estimates of limiting magnitude were performed 
by standard methods, they are in a good agreement with those  according to 
the SIGNAL package created by the team of the Isaac Newton Group of Telescopes 
(http://catserver.ing.iac.es/signal/). The survey speed is simply the field 
area divided by the sum of the exposure time and dead time.

\begin{figure}   % Fig. 4
  \centering
  \includegraphics[width=80mm]{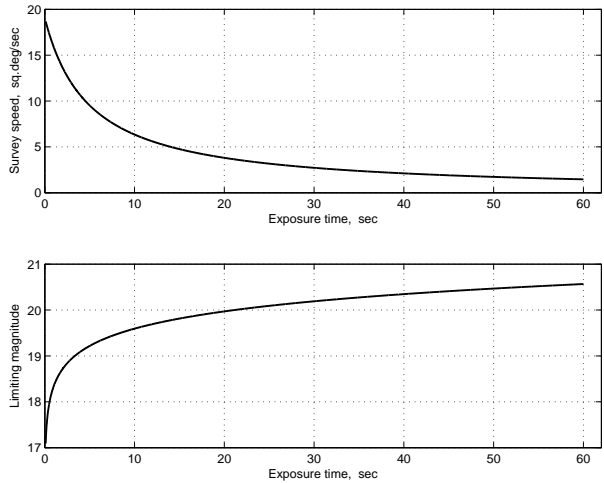} 
  \caption{Survey speed $S$\,(deg$^2$/sec, upper curve) and limiting magnitude
  $m_{lim}$ as the functions of the exposure time (sec) for the VT-119g design.}
\end{figure}

Evidently, the large field of view entails high survey speed of the system. 
It is enough to say that at the exposure time $4$~sec the survey speed of  
$11$\,deg$^2$/sec is attained, so the objects brighter than $\sim 19^m$ can be 
registered in the sky area of $10^4$~deg$^2$ in $15$ minutes. 

As regards limiting magnitude, it is difficult to expect its high value for 
a telescope of relatively small size. We see that for short exposures, which 
are specific to fast transients, the objects not fainter than about $19.5^m$ are 
attainable. The limit is increased to $20^m$ at the exposure time of $20$~sec.
To reach more faint sources, the proposed optical system can be scaled up, 
the more that the image quality is very weakly dependent on the size of the 
system (see the next section).

It seems quite evident that the significant survey efficiency, including both 
the deep limit and high speed, can be achieved by creating a hierarchical system, 
consisting of different types of telescopes.

Let us also estimate the {\em sky survey rate} $\Gamma$, that was defined in 
the review of Terebizh (2011) by equation (A2). It is simply the product of 
the observed sky area $\Omega$ (degree$^2$) and effective aperture area of the 
telescope $A=\pi D_e^2/4$ (meter$^2$), divided by the squared image quality 
$\Delta$ (arc seconds):
\begin{equation}
  \Gamma \equiv \Omega \cdot A/\Delta^2 \quad Herschels,
\end{equation}
where, by definition, the measurement unit is $Herschel \equiv 1\,m^2 
deg^2/arcsec^2$ (shortly denoted by {\em H}). For the fraction of unvignetted 
rays $U=0.90$ and the delivered image quality $\Delta = 3.4''$ we obtain the 
effective aperture diameter $D_e \simeq 0.38$~m and the survey rate 
$\Gamma \simeq 1.5$~{\em H}. Even with such a small part of the available field, 
this is a significant value. 

As was mentioned, detectors may be positioned arbitrarily in the field of view. 
In this connection Tonry (2015) noted, that some optimal summary area of 
detectors to maximize $\Gamma$ should exist, because by expanding the operating 
area we increase $\Omega$, but reduce the effective aperture of the telescope 
$D_e$ due to larger obscuration. Our approximate analytical calculations show 
that this is the case; an accurate assessment can be found for each specific 
task. 

% -------------------- Table 3
\begin{table}
\caption{VT-119f design with an aperture of 800~mm and $30^\circ$ field 
of view. The effective focal length is 1520~mm ($f/1.90$).}
\begin{tabular}{cccccc}
\hline \noalign{\smallskip} 
Surf.& Com-   & $R_0$       & $T$         & Glass   & $D$    \\
 No. & ments  & (mm)        & (mm)        &         & (mm)   \\
\hline 
 1   & L1     & 2411.655    & 94.000      & FS      & 1035.1 \\
 2   &        & 2670.891    & 351.049     & --      & 992.5  \\
 3   & L2     & $-$1955.47  & 104.000     & FS      & 846.7  \\ 
 4   &        & $-$1854.17  & 0.000       & --      & 822.9  \\
 5   & Stop   & $\infty$    & 611.829     & --      & 796.4  \\
 6   & L3     & $-$1005.61  & 100.000     & FS      & 1034.0 \\
 7   &        & $-$854.950  & 54.835      & --      & 1072.4 \\
 8   & L4     & $-$861.958  & 124.000     & FS      & 1090.1 \\
 9   &        & $-$1251.91  & 1988.336    & --      & 1206.7 \\
10   & M1     & $-$3151.84  & $-$1629.20  & Mirror  & 2273.9 \\
11   & Image  & $-$1484.81  & --          & --      & 786.2 \\
\noalign{\smallskip}\hline
\end{tabular} 

Notes to Table~3:\\  
Designations are the same as in Table~1. All surfaces are spheres.
\end{table}

\section{Scaling of the system}

Scaling both up and down leads to attractive systems; we confine ourselves to 
the first option. 

Model VT-119f (Table~3) with the aperture $800$~mm in diameter is a slightly 
optimized doubling of the design VT-119g at the same angular field size. As
before, all lenses are spherical and made of fused silica. It turned out that 
the image quality of both systems, the original and scaled one, is substantially 
the same in the angular measure, $D_{80} \simeq 1.7''$. This means, in particular, 
that for the equal exposure times system VT-119f provides about one magnitude
fainter objects, than VT-119g. (The doubling of diameter gives only $0.75^m$, 
but we get an additional gain due to increased focal length.) 

Since the system under consideration can be essentially scaled up at maintaining 
angular resolution over the whole field, one can get a theoretical design 
of even larger aperture. Unfortunately, not optics by itself, but practical 
limitations on a lens size put the limit on further scaling. In particular, the
diameter of the last lens in the system VT-119f is about $1.2$~m, which is 
already not far from the size reached by the modern technology based on the use 
of glass. Because of soft tolerances, some plastic materials may be promising for 
the lenses. For example, replacement of fused silica in the system VT-119f with 
acrylic reduces the weight of the lens corrector twice at the same image quality
and transparency in a wide spectral range. Perhaps, further research will produce 
the more stable lenses made of plastic materials, especially of a simple 
spherical shape. 

Another way, which is worth studying now, is to use the proposed design
as the core of some light-gathering system of larger size.

\section{Concluding remarks} 

The curvature of the focal surface is as natural for wide-field telescopes, as 
for the human eye. The key step at the transition from conventional telescopes 
to really wide-field systems is changing the type of symmetry of an optical 
system. As noted by G.H. Smith (1998) concerning the Schmidt camera, 
\begin{quotation} 
{{\small There is now point symmetry about the center of the stop (and the 
center of curvature of the mirror), rather than rotational symmetry about an 
axis.}}
\end{quotation} 
However, it must be admitted that the point symmetry is insufficiently perfect 
as long as aspheric surfaces are used, and only application of the all-spherical 
optics makes it strict except inevitable vignetting of the aperture stop. 
Just this feature allowed to expand radically the field of view. It can be shown 
that the proposed here all-spherical system provides the high quality images 
with the field size of more than $50^\circ$.

Perhaps, the Cassegrainian versions of the system are possible only for not too 
wide angular field, say, not larger than $10^\circ$. Behind this approximate 
boundary, it is difficult to account for the curvature of the focal surface.

Of course, large curved light detectors, the production of which has just begun, 
will be used in future. This field is developing rapidly. The principal issues 
and real examples are discussed by Iwert and Delabre (2010), Iwert et al. (2012); 
the first paper includes a photograph of curved detector with size of 60~mm 
$\times$ 60~mm and curvature radius 500~mm. There are also working devices of 
this type. In particular, curved detector has been implemented in the DARPA 
3.5~m Space Surveillance Telescope (Blake et al. 2013). 

Besides, one need to keep in mind the long-known technology based on a plurality 
of delicate waveguides with a curved in aggregate input faced to the focal 
surface. 

The method of working with the spherical focal surface applicable currently is 
the using of small flat detectors each of which is equipped with a flattening lens. 
This way has been implemented, e.g., in the {\em Kepler} space telescope that has 
the $95$~cm aperture and the equivalent field diameter of $11.6^\circ$. Its 
detector consists of 21 pairs of ordinary 59~mm $\times$ 28~mm CCDs covered by 
sapphire field-flattening lenses. As to the VT-119g design, the curvature radius 
of its focal surface is 782~mm, so with a small flat detector of size, say, 25~mm 
the edge images are blurred up to $50$ microns. Our preliminary consideration 
shows that the image quality can be improved considerably even by a single lenslet 
made of fused silica, and is fully recovered by the doublet of the same material. 
The problem becomes much simpler for larger telescopes similar to VT-119f. 
 
It seems likely that further research will expand the scope of the proposed
optical layout. In particular, applications in the spectroscopy, physics of 
cosmic rays, geophysics and tomography are especially promising.

\section*{Acknowledgments}

I thank M.R.~Ackermann (University of New Mexico, U.S.A.), 
M.~Boer (Recherche CNRS ARTEMIS, France), 
R.~Ceragioli (University of Arizona, U.S.A.), 
Yu.A.~Petrunin (Telescope Engineering Company, U.S.A.), 
W.~Stephani (Hamburg University, Germany)
and J.L.~Tonry (Institute for Astronomy, University of Hawaii, U.S.A.) 
for creative discussions. 
Referee proposals were useful.


\begin{thebibliography}{99}

\bibitem{}
Baker, J.G. 1945, U.S. Patent 2,458,132 

\bibitem{}
Baker, J.G. 1962, U.S. Patent 3,022,708 

\bibitem{}
Blake T., Pearce E., Gregory J.A., Smith A., et al., 2013, 
AMOS Technical Conference, 57 

\bibitem{}
Bouwers, A, 1946, Achievements in Optics, Elsevier, Amsterdam, 25 

\bibitem{} 
Busch, W., Ceragioli, R.C., Stephani, W. 2013, Journal of 
Astronomical History and Heritage 16(2), 107 

\bibitem{} 
Carter, B.D., Ashley, M.C.B., Sun, Y-S., Storey, J.W.V. 1992, PASA 10, 74 

\bibitem{}
Hawkins, D.G., Linfoot, E.H. 1945, MNRAS 105, 334 

\bibitem{}
Henize, K.G. 1957, Sky and Telescope 16, 108 

\bibitem{}
Iwert, O., Delabre, B., 2010, Proc. SPIE 7742-27 

\bibitem{}
Iwert, O., Ouelette, D., Lesser M., Delabre, B., 2012, Proc. SPIE 8453-68 

\bibitem{}
Maksutov, D.D. 1944, J. Opt. Soc. Am. 34, 270 

\bibitem{}
Rutten, H.G.J., van Venrooij, M.A.M. 1999, Telescope Optics, 
Willmann-Bell, Ch.~8  

\bibitem{}
Sasaki, M., Kusaka, A., Asaoka, Y. 2002, arXiv:astro-ph/0203348v2 

\bibitem{}
Schmidt, B. 1931, Centr. Ztg. f. Opt. u. Mech., 52, 25 

\bibitem{}
Schroeder, D.J. 2000, Astronomical optics, Academic Press, Ch.~7 

\bibitem{}
Smith, G.H. 1998, Practical Computer-Aided Lens Design,
Willmann-Bell, Richmond, p.~380 

\bibitem{}
Terebizh, V.Yu. 2011. Astron. Nachr. / AN, 332, 714 

\bibitem{} 
Terebizh, V.Yu. 2015, arXiv:1507.07110v1 [astro-ph.IM] 25 Jul 2015 

\bibitem{} 
Tonry, J.L. 2015, Private communication 

\bibitem{}
Wachmann, A.A. 1955, Sky and Telescope 15, No.~1, 4 

\bibitem{}
Wilson, R.N. 1996, Reflective Telescope Optics, Springer, V.~I, Section~3.6 

\bibitem{} 
Wynne, C.G. 1947, MNRAS 107, 356 
	
\end{thebibliography}
\end{document}